\documentclass[notitlepage,a4,article,nofootinbib]{revtex4}

\usepackage{amsmath}%
\usepackage{amsfonts}%
\usepackage{amssymb}%
\usepackage{mathrsfs}
\usepackage{amsthm}
\usepackage{breqn}

\newcommand{\mc}{\mathcal}

\newcommand{\bol}{\boldsymbol}

\newcommand{\abs}[1]{\left\lvert{#1}\right\rvert}

\newcommand{\lr}[1]{\left({#1}\right)}

\newcommand{\mf}{\mathfrak}
\newcommand{\p}{\partial}
\newcommand{\pa}{\partial}

\newcommand{\h}[1]{\hat{#1}}
\newcommand{\mb}[1]{\mathbb{#1}}

\makeatletter
\let\cat@comma@active\@empty
\makeatother

\begin{document}
\title{Superharmonic Instability of Nonlinear Traveling Wave 
Solutions\\ in Hamiltonian systems}
\author{N. Sato and M. Yamada}
\affiliation{Research Institute for Mathematical Sciences, Kyoto University, Kyoto 606-8502, Japan}
\date{\today}

\begin{abstract}
The problem of linear instability of a nonlinear traveling wave 
in a canonical Hamiltonian system with translational symmetry 
subject to superharmonic perturbations is discussed. 
It is shown that exchange of stability occurs when energy is stationary as a function of wave speed.
This generalizes a result proved by \citet{Saffman85} for traveling wave solutions exhibiting 
a wave profile with reflectional symmetry.
The present argument remains true for any noncanonical Hamiltonian system that can be cast in Darboux form, 
i.e. a canonical Hamiltonian form on a submanifold defined by constraints, 
such as a two-dimensional surface wave on a shearing flow, 
revealing a general feature of Hamiltonian dynamics.
\end{abstract}


\maketitle

\begin{normalsize}

\section{Introduction}

Water waves (\citet{Stokes47}) exhibit different types of instabilities,
such as the sideband instability described by \citet{Benjamin67}.
In his study on the instability of finite amplitude water waves, 
\citet{Saffman85} proved a general result regarding the instability of traveling wave solutions
to canonical Hamiltonian systems subject to superharmonic perturbations, i.e. perturbations 
that are periodic in the wave length of the traveling wave. 
He showed that any traveling wave solution would lose stability 
at those wave speeds representing stationary points of the energy
whenever the underlying Hamiltonian system exhibited symmetry 
under translation and reflection of the wave profile.
Here, translational and reflectional symmetry 
are defined as invariance of wave solutions under the exchanges $x'\rightarrow x'+\xi$ and $x'\rightarrow -x'$ respectively, where $\xi\in\mathbb{R}$ is a real constant and  $x'$
the spatial coordinate of a reference frame moving
with the phase speed $\bol{c}$ of the traveling wave 
and with the $x'$-axis oriented along $\bol{c}$.
Saffman's purpose was to explain a numerical result found by \citet{Tanaka83},
who observed the destabilization of a steep gravity wave at the maximum of the energy with respect to wave speed.
Tanaka's observation was unexpected because it conflicted with the conjecture 
made by \citet{Longuet-Higgins78} in the first work on the topic.
Indeed, based on numerical evidence up to the wave steepness $ak=0.42$,
Longuet-Higgins suggested that superharmonic instability would first occur at the wave steepness $ak=0.436$
(a wave height to wavelength ratio of $h/\lambda=0.1388$) corresponding to the maximum wave speed. 
The discrepancy between Tanaka's result and Longuet-Higgins' conjecture was 
due to poor convergence of the numerical scheme adopted by Longuet-Higgins at large steepness.
\citet{Tanaka85} later showed that the eigenvectors associated to the first two modes
of oscillation become linearly dependent at the critical point,  
corroborating bifurcation analysis of \citet{Longuet-Higgins84} and \citet{Chen80}. 
Using Zakharov's Hamiltonian formulation of the water wave problem \citep{Zakharov68,Zakharov97}, 
Saffman performed a linear stability analysis of a traveling wave with reflectionally symmetric profile,
i.e. a profile such that $\bol{A}=\bol{A}^{\ast}$, with $\bol{A}$ and $\bol{A}^{\ast}$ the steady values of the canonical variables of the system in a frame comoving with the wave (see equation (6) of \citet{Saffman85}),   
and analytically proved that the zero eigenvalue arising from 
the translational symmetry of the solutions had algebraic multiplicity
equal or greater than four 
at stationary points of the energy.
In practical terms, this behavior, referred to as exchange of stability (see \citet{Saffman85,Crawford91}), 
indicates the onset of unstable growing modes that eventually destabilize the traveling wave. 
This mechanism can be understood as follows. Due to the Hamiltonian nature of dynamics,
eigenvalues are paired. The zero eigenvalue, which has even algebraic multiplicity equal or greater than two, 
physically correspond to a phase shift of the translationally symmetric fundamental mode of oscillation.
When a purely imaginary eigenvalue crosses the zero eigenvalue at stationary points of energy,  
the square of the purely imaginary eigenvalue undergoes a sign flip.
This change of sign determines the transition of the 
imaginary eigenvalue to the unstable real domain (see e.g. \citet{Tanaka83,Longuet-Higgins78,MacKay86}).
Saffman's Hamiltonian approach for periodic waves in deep water 
was later generalized by \citet{Zufiria86} to include solitary waves on
water of finite depth. 
Most importantly for our purpose, the whole argument remains valid 
for any canonical Hamiltonian system with the aforementioned symmetries.

The aim of the present paper is to show that Saffman's result can be further
generalized to the extent that symmetry under reflection of the wave profile 
is not necessary: 
any translationally symmetric traveling wave solution to a canonical Hamiltonian system 
exhibits superharmonic exchange of stability when energy is stationary with respect to wave speed. 
Notice that, in the present analysis, it is assumed that the only degeneracy of the system is the one 
associated to translational symmetry. Mathematically, this means that the geometric multiplicity of the zero eigenvalue is always unity. 
This working hypothesis is the same used in the analysis of \citet{Saffman85}. 
However, while in \citet{Saffman85} the quadruple multiplicity of the zero eigenvalue
is obtained by direct evaluation of the generalized eigenvectors (see also \citet{Zufiria86}), 
our result relies on the parity of the characteristic polynomial of the 
matrix associated to the linearized equation for the growth of inifinitesimal
perturbations. 
This second approach has the advantage of not requiring 
the assumption of reflectional symmetry $\bol{A}=\bol{A}^{\ast}$, 
therefore allowing the generalization of Saffman's result to asymmetric waves.

In the context of fluid dynamics, this result applies to the superharmonic instability of reflectionally asymmetric water wave profiles \citep{Zufiria87}, waves in multilayered fluids \citep{Constantin15}, and rotational water waves \citep{Constantin07,Wahlen07}. 
Indeed, such systems are not, in general, symmetric under reflection of the wave profile.
However, they arise in canonical Hamiltonian form and admit translationally symmetric traveling wave solutions.
In particular, we expect Tanaka's result to remain true for water waves with constant vorticity 
at their local extrema of energy with respect to the wave velocity.
Indeed, in light of the canonical Hamiltonian formulation of water waves
with constant vorticity due to \citet{Wahlen07}, the present analysis implies
that the same conditions for exchange of stability as the irrotational case hold, even for reflectionally asymmetric wave profiles.  
Reflectionally symmetric water waves with constant vorticity have been observed numerically (see for example \citet{TelesDaSilva88}), and their stability has been examined by \citet{Francius17}. Limited to such symmetric case, the occurrence of Tanaka's instability has been recently demonstrated by \citet{Murashige19} through numerical experiments. 

The present paper is organized as follows. 
In section 2, we review how a continuous canonical Hamiltonian system
can be transformed to an equivalent system in Fourier space
by using a canonical change of variables analogous to that introduced by \citet{Zakharov68}.
In section 3, we follow the linear stability analysis of \citet{Saffman85}, 
without, however, imposing reflectional symmetry, and 
prove that the characteristic polynomial is an even function of the
eigenvalue determining the growth rate of perturbations.
The eigenvector associated to translational symmetry is also obtained.
In section 4, the result of section 3 is used to
show that, in the presence of translational symmetry, 
the algebraic multiplicity of the zero eigenvalue becomes equal
or greater than four 
when energy is stationary with respect to wave speed.
Hence, translationally symmetric traveling wave solutions 
of a canonical Hamiltonian system exchange their stability at the
local extrema of energy.
In section 5, we discuss the occurrence
of Tanaka's instability in water waves on a linear shear flow.
Here, we provide an alternative derivation of the canonical Hamiltonian formulation of water waves with constant vorticity due to Wahlen (2007).
Then, we show that traveling wave solutions exhibit translational symmetry. Hence, according to the result proven in sections 3 and 4, the conditions for exchange of stability to occur at the extrema of the energy are fully verified.
The generalization of the theory to noncanonical Hamiltonian systems
exhibiting traveling wave solutions with translational symmetry is also 
addressed. Finally, conclusions are drawn in section 6.

\section{Preliminaries}
We consider a continuous mechanical system described by a Hamiltonian $H$ and canonical variables $\lr{\zeta,\eta}$, 
\begin{equation}
\zeta_{t}=-\frac{\delta H}{\delta \eta},~~~~\eta_{t}=\frac{\delta H}{\delta\zeta}.\label{Ham1}
\end{equation}  
Here $t$ is the time variable, $H=H\left[\zeta,\eta\right]$ a functional of 
the functions $\zeta\lr{\bol{x},t}$ and $\eta\lr{\bol{x},t}$, with $\bol{x}\in\mathbb{R}^m$ and $m$ a positive integer, and $\delta$ denotes functional derivatives. 
Both the Hamiltonian $H$ and the canonical variables
$\lr{\zeta,\eta}$ take real values. 
In the water wave system, $\zeta$ represents the velocity potential at the water surface and 
$\eta$ the elevation of the wave profile.

The purpose of this section is to obtain a form of system (\ref{Ham1})
in Fourier space that will be useful for the subsequent linear stability analysis. 
The procedure adopted here to transform the continuous system (\ref{Ham1}) 
is analogous to that discussed in \citet{Zakharov68}, \citet{Saffman85}, and \citet{Zufiria86}.

Before carrying out the change to Fourier variables, it is convenient to  introduce the following canonical transformation with complex coefficients,
\begin{equation}
a=\frac{\eta+i\zeta}{\sqrt{2}},~~~~b=\frac{\eta-i\zeta}{\sqrt{2}}.
\end{equation}
Note that $b=a^{\ast}$, since $\lr{\eta,\zeta}$ are real valued.
Here $*$ denotes complex conjugation. 
The inverse transformation is
\begin{equation}
\eta=\frac{a+b}{\sqrt{2}},~~~~\zeta=\frac{a-b}{i\sqrt{2}}.
\end{equation}
In terms of $\lr{a,b}$ system (\ref{Ham1}) reads
\begin{equation}
a_{t}=-i\frac{\delta H}{\delta b},~~~~b_{t}=i\frac{\delta H}{\delta a},\label{Ham2}
\end{equation}
with $H=H\left[a,b\right]$.
Next, we perform the Fourier transform of system (\ref{Ham2}), where the Fourier transform $\mc{F}\left[\theta\right]=\h{\theta}$ of a function $\theta=\theta\lr{\bol{x},t}$ is defined as
\begin{equation}
\h{\theta}\lr{\bol{k},t}=\frac{1}{\lr{2\pi}^{\frac{m}{2}}}\int_{\mb{R}^{m}}{\theta\lr{\bol{x},t}\,e^{-i\bol{k}\cdot\bol{x}}}\,d\bol{x},~~~~\bol{k}\in\mathbb{R}^{m}.\label{FT}
\end{equation}
The inverse Fourier transform is defined accordingly as
\begin{equation}
{\theta}\lr{\bol{x},t}=\frac{1}{\lr{2\pi}^{\frac{m}{2}}}\int_{\mb{R}^{m}}{\h{\theta}\lr{\bol{k},t}\,e^{i\bol{k}\cdot\bol{x}}}\,d\bol{k}.\label{IFT}
\end{equation}
Using equation (\ref{FT}), one has
\begin{equation}
\frac{\delta H}{\delta {\theta}}=\int_{\mathbb{R}^{m}}\frac{\delta H}{\delta\h{\theta}}\frac{\delta\h{\theta}}{\delta{\theta}}\,d\bol{k}=\frac{1}{\lr{2\pi}^{\frac{m}{2}}}\int_{\mathbb{R}^{m}}\frac{\delta H}{\delta\h{\theta}}e^{-i\bol{k}\cdot\bol{x}}\,d\bol{k}. 
\end{equation}
Hence, system (\ref{Ham2}) transforms to
\begin{equation}
\h{a}_{t}\lr{\bol{k}}=-i\frac{\delta H}{\delta \h{b}}\lr{-\bol{k}},~~~~
\h{b}_{t}\lr{\bol{k}}=i\frac{\delta H}{\delta \h{a}}\lr{-\bol{k}}.\label{Ham3}
\end{equation}
Here, the dependence on the time variable has been omitted to simplify the notation.
Next, notice that, from the definition of Fourier transform, the following relations hold:
\begin{equation}
\h{a}\lr{\bol{k}}=\h{b}^{\ast}\lr{-\bol{k}},~~~~\h{b}\lr{\bol{k}}=\h{a}^{\ast}\lr{-\bol{k}}.\label{Ham4_0}
\end{equation}
Therefore, the left-hand side of system (\ref{Ham3}) can be rewritten so that
\begin{equation}
\h{a}_{t}^{\ast}=i\frac{\delta H}{\delta \h{a}},~~~~
\h{b}_{t}^{\ast}=-i\frac{\delta H}{\delta \h{b}}.\label{Ham4_1}
\end{equation}
In this notation, the dependence on the vector $\bol{k}$ has been omitted because redundant.
The same convention will be used in the rest of the paper.
Taking the complex conjugate of (\ref{Ham4_1}) gives 
\begin{equation}
\h{a}_{t}=-i\lr{\frac{\delta H}{\delta \h{a}}}^{\ast},~~~~
\h{b}_{t}=i\lr{\frac{\delta H}{\delta \h{b}}}^{\ast}.\label{Ham4_2}
\end{equation} 

Now consider how equation (\ref{Ham4_1}) is modified by a Galilean transformation
$\lr{\bol{x},t}\mapsto\lr{\bol{x}',t'}=\lr{\bol{x}-\bol{c}\,t,t}$ to a reference frame
moving with constant speed $\bol{c}\in\mathbb{R}^{m}$.
Define $s={a}(\bol{x}^{'}+\bol{c}\,t,t)$. Then,
\begin{equation}
\h{a}=\frac{1}{\lr{2\pi}^{\frac{m}{2}}}\int_{\mb{R}^{m}}{a\,e^{-i\bol{k}\cdot\bol{x}t}}\,d\bol{x}=\frac{e^{-i\bol{k}\cdot\bol{c}t}}{\lr{2\pi}^{\frac{m}{2}}}\int_{\mb{R}^{m}}{s\,e^{-i\bol{k}\cdot\bol{x}^{'}}}\,d\bol{x}^{'}=e^{-i\bol{k}\cdot\bol{c}t}\h{s}.
\end{equation}
In a similar manner, by setting $u=b(\bol{x}^{'}+\bol{c}\,t,t)$, one has $\h{b}=e^{-i\bol{k}\cdot\bol{c}t}\h{u}$. System \ref{Ham4_1} thus becomes
\begin{equation}
\h{s}_{t}^{\ast}=-i\bol{k}\cdot\bol{c}\,\h{s}^{\ast}+i\frac{\delta H}{\delta \h{s}},~~~~
\h{u}_{t}^{\ast}=-i\bol{k}\cdot\bol{c}\,\h{u}^{\ast}-i\frac{\delta H}{\delta \h{u}}.\label{Ham5}
\end{equation}


\section{Superharmonics and linear stability}
In the following we are interested in steady solutions of system (\ref{Ham5}),
i.e. traveling wave solutions of system (\ref{Ham4_1}) 
with wave speed $\bol{c}$ in the reference frame $\lr{\bol{x},t}$. 
Let $\lambda_{0}$ and $k_{0}=2\pi/\lambda_{0}$ denote the wavelength and the wavenumber of the steady wave.
We consider the linear stability of the steady wave with respect 
to superharmonic disturbances, i.e. perturbations with wave vector $\bol{k}^{j}=jk_{0}\bol{c}/\abs{\bol{c}}$ and 
wavelength $\lambda^{j}=\abs{j}^{-1}\lambda_{0}$, $\abs{j}\in\mathbb{N}$. 
Under these conditions, the energy $H$ becomes a discrete sum of
energy contributions corresponding to the modes $k^{j}=jk_{0}$, 
while the variables $\lr{\h{s},\h{u}}$ are now vectors:
\begin{equation}
H=H\lr{\h{s},\h{u}},~~~~\h{s}=\lr{\h{s}^{1},\h{s}^2,...}^{T},~~~~\h{u}=\lr{\h{u}^{1},\h{u}^2,...}^{T},
\end{equation}
with $\h{s}^{j}=\h{s}\lr{\bol{k}^{j}}$, $\h{u}^{j}=\h{u}\lr{\bol{k}^{j}}$, and $T$ indicating the transpose.
Observe that the energy $H$ is now a function of the vectors $\lr{\h{s},\h{u}}$.
Hence, functional derivatives are replaced by partial derivatives, and equation (\ref{Ham5})
becomes
\begin{equation}
\h{s}_{t}^{\ast}=-i\mc{I}\,\h{s}^{\ast}+i\frac{\pa H}{\pa \h{s}},~~~~
\h{u}_{t}^{\ast}=-i\mc{I}\,\h{u}^{\ast}-i\frac{\pa H}{\pa \h{u}}.\label{Ham6}
\end{equation}
Here, $\mc{I}$ is a square diagonal matrix with components $\mc{I}^{i}_{\,j}=cjk_{0}\delta^{i}_{\,j}$ where $\delta^{i}_{j}$ denotes the Kronecker delta, and $c=\abs{\bol{c}}$. 

Let $\lr{S,U}$ identify a steady solution of system (\ref{Ham6}). Such solution must satisfy
\begin{equation}
0=-i\mc{I}\,{S}^{\ast}+i\frac{\pa H}{\pa {S}},~~~~
0=-i\mc{I}\,{U}^{\ast}-i\frac{\pa H}{\pa {U}},\label{Ham7}
\end{equation}
where $\frac{\pa H}{\pa S}=\lr{\frac{\pa H}{\pa \h{s}}}_{\lr{\h{s},\h{u}}=\lr{S,U}}$ and $\frac{\pa H}{\pa U}=\lr{\frac{\pa H}{\pa \h{u}}}_{\lr{\h{s},\h{u}}=\lr{S,U}}$. This notation will be used throughout the paper.
Notice that now the discretized equations \eqref{Ham6} and \eqref{Ham7}
correspond to equations (2) and (3) of \citet{Saffman85}.
However, observe that, in contrast with equation (6) of \citet{Saffman85},
which implies reflectional symmetry of the wave profile (see the discussion after equations (15), (30), and (31) in \cite{Zufiria86} on this point), 
such symmetry is not assumed here.
In practice, the absence of reflectional symmetry makes the Jacobian
matrix $J$ (to be defined shortly) of the linearized equation 
for the evolution of perturbations complex valued.
For this reason, the proof that the zero eigenvalue has at least quadruple multiplicity requires a different approach based on the parity of the characteristic polynomial associated to $J$, which is a property that can be deduced from the Hamiltonian nature of the equations.  

Next, consider infinitesimal perturbations $\lr{d,f}$ of the equilibrium state $\lr{S,U}$,
\begin{equation}
\h{s}=S+\epsilon d,~~~~\h{u}=U+\epsilon f,\label{delta}
\end{equation}
with $\epsilon$ a small positive real constant. 
By substituting \eqref{delta} into \eqref{Ham7} and by linearizing
around the equilibrium state with respect to the small parameter $\epsilon$, 
one arrives at the system of linear equations
\begin{equation}
d_{t}^{\ast}=-i\mc{I}d^{\ast}+iMf+iNd,~~~~
f_{t}^{\ast}=-i\mc{I}f^{\ast}-iM^{T}d-iPf,\label{delta2}
\end{equation}
with $M=\frac{\p^2 H}{\p S\p U}$, $N=\frac{\p^2 H}{\p S^2}$, and $P=\frac{\p^2 H}{\p U^2}$.
Note that the matrices $N$ and $P$ are symmetric.
It is convenient to put equation (\ref{delta2}) in matrix form.
Define the vector
\begin{equation}
v=\lr{d,f,d^{\ast},f^{\ast}}^{T},
\end{equation}
and the matrix
\begin{equation}
J=i\begin{bmatrix}
\mc{I}&0 &-N^{\ast} &-M^{\ast} \\
0&\mc{I}& M^{\dagger} & P^{\ast}\\
N & M & -\mc{I} & 0 \\
-M^{T} & -P & 0 &-\mc{I}.
\end{bmatrix}
\end{equation}
Here, $M^{\dagger}$ denotes the transpose conjugate of $M$.
Then, the linearized system (\ref{delta2}) can be written as
\begin{equation}
v_{t}=Jv.\label{delta3}
\end{equation}
The growth rate of infinitesimal perturbations $v$ can thus
be determined by solving the eigenvalue problem for the matrix $J$.

Owing to 
the Hamiltonian nature of the original system of equations, equation (\ref{Ham1}),  
the eigenvalues $\sigma$ of the matrix $J$ come in 
pairs $\pm\sigma$. 
If $d^{j}_{+}=d_{0}^{j}e^{i\sigma t}$, with $d_{0}^{j}=d_{0}(k^{j})$ a function of $k^{j}$, is a solution, 
so must be $d^{j}_{-}=d_{0}^{j}e^{-i\sigma t}$.
To see this explicitly, consider the following matrix:
\begin{equation}
Q=\begin{bmatrix} 0& 0& -I_{0}& 0\\ 0& 0& 0& I_{0}\\ I_{0}& 0& 0& 0\\ 0& -I_{0}& 0& 0\end{bmatrix}.
\end{equation}
Here, $I_{0}$ is the identity matrix with the same dimension as that of $M$, $N$, and $P$. 
We have $Q^2=-I$, where $I$ is the identity matrix with the same dimension as that of $Q$, and $Q^{T}=-Q$.
Using the fact that the matrices $N$ and $P$ are symmetric, 
one can verify the following identity
\begin{equation}
\lr{QJ}^{T}=QJ.
\end{equation}
It follows that 
\begin{equation}
Q{J}^{T}Q^{T}=-J.
\end{equation}
Then, the characteristic polynomial $p\lr{\sigma}$ of the matrix $J$ satisfies
\begin{equation}
p\lr{\sigma}=\det\lr{J-\sigma I}=\det\lr{-QJ^{T}Q^{T}+\sigma Q^2}=\det\lr{Q}^2\det\lr{J^{T}+\sigma I}=p\lr{-\sigma}.
\end{equation}  
We have thus shown that the characteristic polynomial is an even function of $\sigma$.
This result implies that eigenvalues come in pairs $\pm\sigma$, 
and that any zero eigenvalue $\sigma=0$ must have
even algebraic multiplicity equal or greater than two. 

$\sigma=0$ is an eigenvalue of the matrix $J$. Indeed, by hypothesis
solutions of system (\ref{Ham1}) are invariant under translations. 
Steady solutions of system (\ref{delta3}) are therefore not unique, and the kernel
of the matrix $J$ must be non-empty. This implies the existence of an eigenvector $\alpha$
with eigenvalue $\sigma=0$. 
Notice that it is assumed here that the eigenvector $\alpha$ associated to invariance under translation 
is the only degeneracy of the matrix $J$.  
The eigenvector $\alpha$ can be evaluated by observing that if $S^{j}=S\lr{jk_{0}}$
is a steady solution, so must be $S^{j}_{\xi}=e^{ij\xi}S^{j}$ for any choice of the displacement $\xi\in\mathbb{R}$.
Then, the eigenvector $\alpha$ will be the infinitesimal generator of the Lie group associated to translation of
solutions,
\begin{equation}
\alpha=-
\begin{bmatrix}\lr{\frac{d{S}_{\xi}}{d\xi}}_{\xi=0}\\
\lr{\frac{d{U}_{\xi}}{d\xi}}_{\xi=0} \\ 
\lr{\frac{d{S}^{\ast}_{\xi}}{d\xi}}_{\xi=0} \\
\lr{\frac{d{U}^{\ast}_{\xi}}{d\xi}}_{\xi=0}
\end{bmatrix}=
\frac{i}{ck_{0}}\begin{bmatrix}
-\mc{I}S\\
-\mc{I}U\\
\mc{I}S^{\ast}\\
\mc{I}U^{\ast}
\end{bmatrix}.
\end{equation}
Here we used the fact that translating $U\lr{\bol{k}}$ is tantamount to translating $S^{\ast}\lr{-\bol{k}}$. 
The minus sign in the definition of $\alpha$ was put for later convenience.

\section{Exchange of stability}
Purpose of the present section is to obtain the conditions under which exchange of stability occurs.
Mathematically, this amounts at determining the values of the wave speed $c$ at which
the algebraic multiplicity of the eigenvalue $\sigma=0$ is greater than two.
In particular, it will be shown that this occurs at stationary points of the energy, $\frac{dH}{dc}=0$,
and that, at such points, the algebraic multiplicity is equal or greater than four, while the geometric multiplicity is unity. In contrast with \citet{Saffman85}, symmetry under reflection of the wave
profile is not used in the analysis, thereby generalizing his result.  

First, observe that, from equation (\ref{Ham7}), the variables $\lr{S,U}$, and consequently the Hamiltonian $H\lr{S,U}$ can be regarded as functions of wave speed $c$. 
The energy $H$ is stationary with respect to wave speed $c$ when
\begin{equation}
\frac{dH}{dc}=\frac{\p H}{\p S^{j}}\frac{dS^{j}}{dc}+\frac{\p H}{\p U^{j}}\frac{dU^{j}}{dc}=0.\label{dHdc1}
\end{equation}
In light of (\ref{Ham7}), equation (\ref{dHdc1}) can be written as
\begin{equation}
\frac{dH}{dc}=\lr{\frac{dS}{dc},\mc{I}S}-\lr{\frac{dU}{dc},\mc{I}U}=0,\label{dHdc2}
\end{equation}
where $\lr{x,y}=x\cdot y^{\ast}$ denotes the inner product of two complex vectors $x$ and $y$. 

At this point it is useful to make some considerations on the simplification of equation (\ref{dHdc2}) 
that occurs if the wave profile is symmetric under reflection. 
For this purpose it is sufficient to consider the one-dimensional case $\eta=\eta\lr{x,t}$, with 
$x\in\mathbb{R}$. Symmetry under reflection translates into the condition that
\begin{equation}
\eta\lr{x',t}=\eta\lr{-x',t},~~~~x{'}=x-c\,t.\label{RS}
\end{equation}
On the other hand, recall that $\sqrt{2}\,\eta=s+s^{\ast}$.
In the superharmonic setting, the variable $s$
can be expanded into Fourier series as $s=\sum_{j=-\infty}^{+\infty}\h{s}^{j}e^{ijk_{0}x{'}}$.
Hence, equation (\ref{RS}) implies that
\begin{equation}
0=\sum_{j=-\infty}^{+\infty}\lr{\h{s}^{j}-\h{s}^{j\ast}}\sin\lr{jk_{0}x'}.
\end{equation}
This condition can be satisfied by demanding that $\h{s}=\h{s}^{\ast}$.
In such case, $\h{u}^{-j}=\h{s}^{j\ast}=\h{s}^{j}$.
Furthermore, the Hamiltonian of the water wave system, which can be written
as a function of $\h{s}$ and $\h{s}^{\ast}$, is endowed with
the symmetry $H\lr{\h{s},\h{s}^{\ast}}=H\lr{\h{s}^{\ast},\h{s}}$.
Therefore, if the wave profile is assumed to be symmetric under reflection, 
equation (\ref{dHdc2}) reduces to the requirement
\begin{equation}
0=2\frac{dS}{dc}\cdot\mc{I}S.
\end{equation}
This is the scenario examined by \citet{Saffman85}.

We now return to the problem without reflectional symmetry.
By differentiating the equilibrium system (\ref{Ham7}) with respect to wave velocity $c$,
the following relationships can be derived:
\begin{equation}
\begin{split}
\mc{I}\frac{dS^{\ast}}{dc}-M\frac{dU}{dc}-N\frac{dS}{dc}&=-c^{-1}\mc{I}S^{\ast},\\
\mc{I}\frac{dU^{\ast}}{dc}+P\frac{dU}{dc}+M^{T}\frac{dS}{dc}&=-c^{-1}\mc{I}U^{\ast}.\label{Ham8}
\end{split}
\end{equation}
We further define
\begin{equation}
\beta=\frac{1}{k_{0}}\lr{\frac{dS}{dc},\frac{dU}{dc},\frac{dS^{\ast}}{dc},\frac{dU^{\ast}}{dc}}^{T}.
\end{equation}
Then, system (\ref{Ham8}) has the form
\begin{equation}
J\beta=\alpha,
\end{equation}
implying
\begin{equation}
J^2\beta=0.
\end{equation}
Hence, $\beta$ is a generalized eigenvector of rank two associated to the eigenvalue $\sigma=0$.
This is consistent with the fact that the algebraic multiplicity of $\sigma=0$ is at least two.
Exchange of stability occurs when the eigenvalue $\sigma=0$ has algebraic multiplicity
greater or equal to four, while the geometric multiplicity of the associated eigenvector remains unity.
Since it has been shown that the characteristic polynomial of the matrix $J$ is even,
it is sufficient to determine the values of $c$ such that the algebraic multiplicity of $\sigma=0$
is three to obtain exchange of stability.
 
Let $J^{\dagger}$ denote the transpose conjugate of $J$, 
and ${\rm ker} J^{\dagger}$ the kernel of the operator $J^{\dagger}$. 
We wish to show that, if a vector $f$ is orthogonal to the elements of ${\rm ker}J^{\dagger}$,
i.e. $\lr{f,\alpha^{\dagger}}=0$ $\forall \alpha^{\dagger}\in{\rm ker} J^{\dagger}$,
then $f\in{\rm Im} J$, where ${\rm Im}J$ denotes the image of the operator $J$.
A vector $f$ belongs to ${\rm Im}J$ if there exists a vector $\gamma$ such that $f=J\gamma$.
Suppose that $f\notin{\rm Im}J$. Then $f$ can be decomposed as $f=J\gamma+\epsilon$, with $\epsilon\perp{\rm Im}J$.
Since $f$ is orthogonal to ${\rm ker}J^{\dagger}$, one has 
\begin{equation}
\lr{f,\alpha^{\dagger}}=(J\gamma+\epsilon,\alpha^{\dagger})=\lr{\epsilon,\alpha^{\dagger}}=0~~~~\forall\alpha^{\dagger}\in{\rm ker}J^{\dagger}.
\end{equation}
Hence, $\epsilon\notin{\rm ker}J^{\dagger}$.
On the other hand, $\epsilon$ is, by construction, orthogonal to ${\rm Im}J$, implying that
\begin{equation}
\lr{\epsilon,J\delta}=\lr{J^{\dagger}\epsilon,\delta}=0~~~~\forall\delta.
\end{equation}
But then $\epsilon\in{\rm ker}J^{\dagger}$. Therefore $\epsilon=0$ and $f\in{\rm Im}J$.
By the identification $f=\beta$, it follows that the sufficient condition for 
$\beta$ to belong to the image of $J$ is that
\begin{equation}
\lr{\beta,\alpha^{\dagger}}=0.\label{bad}
\end{equation}  
If such condition is satisfied, $\beta=J\gamma$ for some vector $\gamma$, implying
that $\gamma$ is a generalized eigenvector of rank three associated to the eigenvalue $\sigma=0$.
Due to the parity of the characteristic polynomial, satisfying equation (\ref{bad})
is thus enough to prove that the zero eigenvalue has algebraic multiplicity equal or greater than four and
geometric multiplicity of one.

$\alpha^{\dagger}$ can be evaluated explicitly by demanding that $J^{\dagger}\alpha^{\dagger}=0$ 
and using the condition $J\alpha=0$. The expression of $\alpha^{\dagger}$ is
\begin{equation}
\alpha^{\dagger}=k_{0}\lr{-\mc{I}S,\mc{I}U,-\mc{I}S^{\ast},\mc{I}U^{\ast}}^{T}.
\end{equation}
Equation (\ref{bad}) thus becomes
\begin{equation}
\lr{\beta,\alpha^{\dagger}}=-\lr{\frac{dS}{dc},\mc{I}S}+\lr{\frac{dU}{dc},\mc{I}U}
-\lr{\frac{dS^{\ast}}{dc},\mc{I}S^{\ast}}+\lr{\frac{dU^{\ast}}{dc},\mc{I}U^{\ast}}=0.
\end{equation}
On the other hand, recalling the expression of $dH/dc$, equation (\ref{dHdc2}),
it follows that
\begin{equation}
\lr{\beta,\alpha^{\dagger}}=-2{\rm Re}\lr{\frac{dH}{dc}}=-2\frac{dH}{dc}=0.
\end{equation}
Hence, exchange of stability occurs at stationary points of the energy.


\section{Generalizations and examples: water waves with constant vorticity}

The analysis of the previous sections can be slightly generalized.
First, observe that the number of canonical pairs $\lr{\zeta,\eta}$
does not need to be one: by setting
$\lr{{\zeta},\eta}=\lr{\zeta^{1},\zeta^{2},...,\eta^{1},\eta^{2},...}$,
with $\zeta^{i}\lr{\bol{x},t}$ and  $\eta^{i}\lr{\bol{x},t}$ the $i$th canonical pair,
one can verify that the results of the previous sections remain unchanged.
In particular, they apply to noncanonical Hamiltonian systems
that can be cast in Darboux form, which is a canonical Hamiltonian form on a submanifold 
defined by the constraints (Casimir invariants) that characterize the noncanonical Hamiltonian structure. 
More precisely, consider a noncanonical Hamiltonian system 
\begin{equation}
{\bol{z}}_{t}=\mc{J}\delta_{\bol{z}}H,\label{NHS}
\end{equation}
where $\bol{z}$ takes values in a domain $\Omega\subset\mathbb{R}^{n}$ and  
is an element of a Hilbert space $X$ defining phase space,  
$H\left[\bol{z}\right]$ a smooth function in $X$ representing the Hamiltonian of the system, 
$\delta_{\bol{z}}$ the functional derivative with respect to $\bol{z}$, and $\mc{J}$ the Poisson operator. 
Here, the Poisson operator is defined by the properties of the associated Poisson bracket, 
the bilinear form $\left\{f,g\right\}=\langle \delta_{\bol{z}}f,\mc{J}\delta_{\bol{z}}g \rangle=\int_{\Omega}\frac{\delta f}{\delta z^{i}}\mc{J}^{ij}\frac{\delta g}{\delta z^{j}}\,d\bol{x}$, which
satisfies antisymmetry $\left\{f,g\right\}=-\left\{g,f\right\}$ and Jacobi identity
$\left\{f,\left\{g,h\right\}\right\}+\left\{g,\left\{h,f\right\}\right\}+\left\{h,\left\{f,g\right\}\right\}=0$, with $f$, $g$, and $h$ arbitrary smooth functionals on $X$.
Suppose that there exists a change of variables $\bol{z}\mapsto\lr{\zeta,\eta,C}$
such that the variables $\lr{\zeta,\eta}$ are canonically conjugated, i.e. $\zeta_{t}^{i}=-\frac{\delta H}{\delta\eta^{i}}$ and $\eta_{t}^{i}=\frac{\delta H}{\delta\zeta^{i}}$, and the 
variables $C=\lr{C^{1},C^{2},...}$ are constants of motion (Casimir invariants) due to the property that $\left\{f,C^{j}\right\}=0$ for all smooth functional $f$, which implies $C_{t}^{j}=\left\{C^{j},H\right\}=0$.
Then, the transformed Poisson operator $\mf{J}$ is in Darboux form,
meaning that it acts as a degenerate symplectic matrix, with the
degeneracy represented by the existence of Casimir invariants $C$,
\begin{equation}
\mf{J}=\begin{bmatrix}
0_{m} & -I_{m} & 0 \\
I_{m}& 0_{m} & 0 \\
0 & 0 & 0_{q} 
\end{bmatrix}.
\end{equation}
Here, $0_{m}$ and $I_{m}$ are the $m$-dimensional null and identity matrix respectively, with $m$ the number of
canonical pairs, and $0_{q}$ the $q$-dimensional null matrix, with $q$ the number of Casimir invariants.
If one is able to find such change of variables,
the stability analysis of the present paper applies.
It is worth observing that, however, the existence
of the Darboux form for the Poisson operator is guaranteed only
locally and for finite dimensional systems (Darboux theorem, see \citet{deLeon89}).

\subsection{Water waves with constant vorticity}

As an application of the theory to fluid dynamics, 
we now consider the superharmonic instability of
gravity-capillary waves over a linear shear flow.
The problem can be studied in two dimensions $\lr{x,y}$, with
the horizontal axis $x$ aligned with phase speed, $\bol{c}=c\,\p_{x}$,
and the variable $y$ representing the spatial coordinate along the upward vertical axis.
In this notation, $\p_{x}$ and $\p_{y}$ represent the unit tangent vectors
along the $x$-axis and $y$-axis respectively.
We assume the system to be periodic in the $x$-direction with period $L$. 
The domain occupied by the fluid is $\Omega=\left\{\lr{x,y}\in\mathbb{R}^2:0< x< L, d< y<\eta\lr{x,t}\right\}$ with boundary $\p\Omega=\Sigma \cup B \cup T$, where $\Sigma=\left\{\lr{x,y}\in\mathbb{R}^2:y=\eta\lr{x,t}\right\}$ is the fluid surface, $B=\left\{\lr{x,y}\in\mathbb{R}^2:y=d\right\}$ the fluid bottom
at constant depth $d\leq0$, $T=\left\{\lr{x,y}\in\mathbb{R}^2:x=\left\{0,L\right\}\right\}$ the vertical boundary, and $\eta\lr{x,t}$ the elevation of the water surface at $\lr{x,t}$.
We further demand that the fluid velocity $\bol{v}=\lr{v_{x},v_{y}}$
is given by
\begin{equation}
\bol{v}=\nabla\phi-\omega\, y\,\p_{x}.
\end{equation}
Here, $\omega\in\mathbb{R}$ is a real constant representing the vorticity of the system:
\begin{equation}
\omega=\frac{\p v_{y}}{\p x}-\frac{\p v_{x}}{\p y},
\end{equation}
and the function $\phi$ is the velocity potential.
Then, it can be shown that the two-dimensional ideal Euler equations 
at constant fluid density reduce to the following system of partial
differential equations for the variable $\phi$
\begin{equation}
\phi_{t}=f-\omega\psi-\frac{\abs{\nabla\phi}^2}{2}+\omega y\phi_{x}-P-gy,~~~~\Delta\phi=0~~~~{\rm in}~~\Omega.\label{WWCV}
\end{equation}
Here, $f=f\lr{t}$ is an arbitrary function of time $t$, $P$ the pressure, $g$ the gravitational constant, and the function $\psi$ the harmonic conjugate of $\phi$ such that $\phi_{x}=\psi_{y}$ and $\phi_{y}=-\psi_{x}$ (a lower index indicates derivation; this notation is used in the rest of the paper).   
The function $\phi$ is harmonic as a direct consequence of the continuity equation, which implies that the fluid velocity must be divergence free in a regime of constant fluid density. 
System (\ref{WWCV}) must be further supplied with boundary conditions at the
top surface $\Sigma$, the bottom $B$, and the vertical boundary $T$. 
At the vertical boundary $T$, we impose periodic boundary conditions 
for the velocity, $\bol{v}\lr{0,y,t}=\bol{v}\lr{L,y,t}$. 
At the bottom $B$, we require the Neumann boundary condition $\phi_{y}=0$.
Finally, at the water surface $\Sigma$ we have:
\begin{subequations}\label{WWCVB}
\begin{align}
\phi_{t}&=f-\omega\psi-\frac{\abs{\nabla\phi}^2}{2}+\omega y\phi_{x}-gy+\alpha\frac{\p}{\p x}\lr{\frac{\eta_{x}}{\sqrt{1+\eta_{x}^2}}},\label{ST}\\
\eta_{t}&=\phi_{y}-\phi_{x}\eta_{x}+\omega y\eta_{x}.\label{DBC}
\end{align}
\end{subequations}
Here, the pressure term $P$ was reabsorbed in the definition of $f$, since, at the water surface, $P$ equals the constant atmospheric pressure of the overlying air. 
The last term in equation (\ref{ST}) represents surface tension, with $\alpha\in\mathbb{R}$ a real constant. 
Equation (\ref{DBC}) can be deduced by the fact that 
the quantity $\eta\lr{x,t}-y$ must be preserved along the flow $\bol{v}$.
Observe that a solution $\lr{\phi,\eta}$ of system (\ref{WWCVB}) at the surface $\Sigma$ determines $\phi$ in the whole $\Omega$. Indeed, the value of $\phi$ in the interior domain can be obtained as the unique solution of the second equation of system (\ref{WWCV}) by imposition of the computed dynamic boundary conditions on $\p\Omega$. 
In what follows, we therefore restrict our attention to system (\ref{WWCVB}), 
which represents water waves with constant vorticity.
It is now convenient to introduce the following notation:
\begin{equation}
\vartheta=\phi\rvert_{y=\eta},~~~~\chi=\psi\rvert_{y=\eta}.
\end{equation}

For exchange of stability to occur at the extrema of energy
with respect to wave speed (provided that such extrema exist), 
it is sufficient to show that system (\ref{WWCVB})
has a canonical Hamiltonian form with translationally symmetric
traveling wave solutions.
\citet{Wahlen07} has obtained the canonical Hamiltonian formulation of 
two-dimensional periodic gravity-capillary waves with constant vorticity in water of finite depth by finding the change of variables $\bol{z}=\lr{\vartheta,\eta}\mapsto\lr{\zeta,\eta}$
that transforms the noncanonical Poisson operator $\mc{J}$ 
\begin{equation}
\mc{J}=\begin{bmatrix} \omega\int &-1\\ 1&0\\\end{bmatrix}.\label{NPO}
\end{equation}
of the system into the canonical symplectic matrix
\begin{equation}
\mf{J}=\begin{bmatrix} 0&-1\\ 1&0\\\end{bmatrix}.
\end{equation}
We refer the reader to \citet{Wahlen07} for the precise
definition of the integral operator $\omega\int$ appearing 
in the noncanonical Poisson operator (\ref{NPO}).
Note that, as in the case of irrotational water waves, there is only
a single canonical pair $\lr{\zeta,\eta}=\lr{\zeta^{1},\eta^{1}}$.
Hence, when written in terms of the canonical variables, equation (\ref{NHS})
reduces to the canonical form (\ref{Ham1}):
\begin{equation} 
\begin{bmatrix}\zeta_{t}\\\eta_{t}\end{bmatrix}=\mf{J}\begin{bmatrix}\frac{\delta H}{\delta\zeta}\\\frac{\delta H}{\delta\eta}\end{bmatrix}=\begin{bmatrix}-\frac{\delta H}{\delta\eta}\\\frac{\delta H}{\delta \zeta}\end{bmatrix}.
\end{equation}

In the remaining part of this section, we first 
provide an alternative derivation of the canonical Hamiltonian formulation of water waves with constant vorticity due to \citet{Wahlen07}. 
In \citet{Wahlen07} the form of the canonical momentum $\zeta$ is assumed, and it is shown that, with this choice, the noncanonical Poisson operator transforms to the canonical symplectic matrix. 
Here, we solve Hamilton's canonical equations for the unknown canonical momentum $\zeta$, and verify the obtained result by directly variating the Hamiltonian to produce the equations of motion.
Then, we show that, if traveling wave solutions exist,
they possess translational symmetry. 
Hence, according to the result proven in sections 3 and 4, the conditions for exchange of stability to occur at the extrema of the energy are fully verified. 
Note that there is no need to assume, at any point of the analysis, reflectional symmetry of the wave profile.
We limit our attention to gravity-capillary waves in deep water, i.e. $d\rightarrow-\infty$ (the calculations remain unchanged for other values of $d$).

\subsection{Derivation of the canonical momentum}

The purpose of this subsection is to derive the canonical momentum $\zeta$
conjugated to the elevation $\eta$. 
If $\zeta$ and $\eta$ are canonically conjugated variables, the evolution of $\eta$ must be given by the gradient of the Hamiltonian $H\left[\zeta,\eta\right]$ with respect to $\zeta$. 
In the following calculations it is convenient to adopt the notation $\lr{\zeta,\eta'}$, with $\eta'=\eta$, to avoid ambiguity in the differentation of the Hamiltonian.  
By application of the chain rule for the change of variables $\lr{\zeta,\eta'}\mapsto\lr{\vartheta,\eta}$, we thus have 
\begin{equation}
\eta'_{t}=\frac{\delta H}{\delta\zeta}=\int\lr{\frac{\delta H}{\delta\eta}\frac{\delta\eta}{\delta\zeta}+\frac{\delta H}{\delta\vartheta}\frac{\delta\vartheta}{\delta\zeta}}\,ds=\int\frac{\delta H}{\delta\vartheta}\frac{\delta\vartheta}{\delta\zeta}\,ds=\int\eta_{t}\frac{\delta\vartheta}{\delta\zeta}\,ds.\label{etat_can}
\end{equation}
Here, integration with respect to the variable $s$ is performed on the interval $\left[0,L\right]$, $\eta=\eta\lr{s,t}$, $\vartheta=\vartheta\lr{s,t}$, and $\zeta=\zeta\lr{x,t}$. 
In the last passage, we used the fact that, from equation (\ref{NPO}), $\eta_{t}=\frac{\delta H}{\delta\vartheta}$. Equation (\ref{etat_can}) thus implies that
$\frac{\delta\vartheta}{\delta\zeta}=\delta\lr{x-s}$. Therefore, 
\begin{equation}
\vartheta=\zeta+\sigma\left[\eta'\right],\label{vartheta}
\end{equation}
where $\sigma\left[\eta'\right]$ is an integro-differential operator acting on $\eta'$ to be determined. 
The operator $\sigma$ can be obtained by requiring that the evolution of $\zeta$ is minus the gradient of the Hamiltonian with respect to $\eta'$. Applying the chain rule again, 
\begin{equation}
\zeta_{t}=-\frac{\delta H}{\delta\eta'}=
-\int\lr{\frac{\delta H}{\delta\eta}\frac{\delta\eta}{\delta\eta'}+\frac{\delta H}{\delta\vartheta}\frac{\delta\vartheta}{\delta\eta'}}\,ds=-\frac{\delta H}{\delta\eta}-\int\lr{\eta_{t}\frac{\delta\sigma}{\delta\eta'}}\,ds.\label{zetat_1_}
\end{equation}
In the last passage, we used the fact that $\frac{\delta\eta}{\delta\eta'}=\delta\lr{x-s}$ and $\eta_{t}=\frac{\delta H}{\delta\vartheta}$.
On the other hand, recalling equation (\ref{NPO}), we have
\begin{equation}
\zeta_{t}=\vartheta_{t}-\sigma_{t}=-\frac{\delta H}{\delta\eta}+\omega\int\frac{\delta H}{\delta\vartheta}-\sigma_{t}=-\frac{\delta H}{\delta \eta}+\omega\int\eta_{t}-\sigma_{t}.\label{zetat_2_}
\end{equation}
Here, $\omega\int=\omega\int_{0}^{x}\,ds$ is the integral operator introduced in (\ref{NPO}). Furthermore, setting $\tilde{\chi}=\chi-\frac{\omega}{2}\eta^2$, observe that
\begin{equation}
\begin{split}
-\tilde{\chi}_{x}=\frac{\p}{\p x}\lr{\frac{\omega}{2}\eta^2-\chi}&=\omega\eta\eta_{x}-\psi_{x}\lr{x,\eta,t}-\psi_{y}\lr{x,\eta,t}\eta_{x}\\&=\omega\eta\eta_{x}+\phi_{y}\lr{x,\eta,t}-\phi_{x}\lr{x,\eta,t}\eta_{x}=\eta_{t}.\label{eta_x}
\end{split}
\end{equation}
Hence, substituting (\ref{eta_x}) in (\ref{zetat_1_}) and (\ref{zetat_2_}), and comparing (\ref{zetat_1_}) with (\ref{zetat_2_}), it follows that $\sigma$ must satisfy the condition
\begin{equation}
\lr{\tilde{\chi}\frac{\delta\sigma}{\delta\eta'}}\bigg\rvert_{0}^{L}-\int\lr{\tilde{\chi}\frac{\delta\sigma_{s}}{\delta\eta'}}\,ds=-\omega\tilde{\chi}+\omega\tilde{\chi}\lr{0,t}-\sigma_{t}.\label{sigma}
\end{equation}
Here, the left-hand side was obtained with integration by parts. 
Since $\psi$, and therefore $\tilde{\chi}$, are defined up to an arbitrary function of time, it is convenient to eliminate the time-dependent term $\tilde{\chi}\lr{0,t}$ by redefining $\tilde{\chi}$ as $\tilde{\chi}'=\tilde{\chi}-\tilde{\chi}\lr{0,t}$.
Hence, from conservation of total mass $\frac{d}{d t}\int_{0}^{L}\int_{-\infty}^{\eta}\,dx\,dy=\int_{0}^{L}\eta_{t}\,dx=\tilde{\chi}\lr{0,t}-\tilde{\chi}\lr{L,t}=0$, we obtain $\tilde{\chi}'\lr{0,t}=\tilde{\chi}'\lr{L,t}=0$.
Then, one sees that a solution of equation (\ref{sigma}) is
\begin{equation}
\sigma=\frac{\omega}{2}\int_{0}^{x}\eta\lr{s,t}\,ds.
\end{equation}
From equation (\ref{vartheta}), we conclude that the canonical momentum $\zeta$ is given by
\begin{equation}
\zeta=\vartheta-\frac{\omega}{2}\int_{0}^{x}\eta\lr{s,t}\,ds.
\end{equation}

\subsection{Variation of the Hamiltonian}

The Hamiltonian of the system, representing total kinetic plus potential energy, can be written as
\begin{equation}
H\left[\zeta,\eta\right]=\int_{0}^{L}\left[\int_{-\infty}^{\eta}\lr{\frac{\abs{\nabla\phi}^2}{2}-\omega y\phi_{x}}\,dy\right]\,dx+\int_{0}^{L}\lr{\frac{g}{2}\eta^2+\frac{\omega^2}{6}\eta^3+\alpha\sqrt{1+\eta^{2}_{x}}}\,dx.
\end{equation}
All variations are assumed to vanish at the vertical boundary $T$. 
First, we consider the variation of $H$ with respect to $\zeta$. 
The only term of interest is
\begin{equation}
\delta H=\int_{0}^{L}\int_{-\infty}^{\eta}\nabla\cdot\left[\delta\phi\lr{\nabla\phi-\omega y\nabla x}\right]\,dx\,dy=\int_{0}^{L}\left[\delta\phi\lr{\phi_{y}-\phi_{x}\eta_{x}+\omega\eta\eta_{x}}\right]_{y=\eta}\,dx.
\end{equation}
Here, we used $\Delta\phi=0$ in $\Omega$, the vanishing of variations on $T$, 
and the boundary condition $\phi_{y}=0$ at the bottom.
Now observe that, at fixed $\eta$, $\delta\zeta=\delta\vartheta=\delta\phi\lr{x,\eta,t}$. Hence, recalling equation (\ref{DBC}), at $y=\eta$ we obtain
\begin{equation}
\eta_{t}=\frac{\delta H}{\delta\zeta}=\phi_{y}-\phi_{x}\eta_{x}+\omega\eta\eta_{x}.\label{etat_2}
\end{equation}
Next, we consider variations with respect to $\eta$. At fixed $\zeta$,  
one has
\begin{equation}
0=\delta\zeta=\delta\vartheta-\frac{\omega}{2}\int_{0}^{x}\delta\eta\,ds.\label{dxiatz}
\end{equation}
Suppose that $\phi_{\ast}$ is the solution
of the boundary value problem associated to the change $\delta\eta$. 
Then, from equation (\ref{dxiatz}), we must have
\begin{equation}
\phi_{\ast}\lr{x,\eta+\delta\eta,t}=\vartheta+\delta\vartheta=\phi\lr{x,\eta,t}+\frac{\omega}{2}\int_{0}^{x}\delta\eta\,ds.
\end{equation}
Hence, at first order in $\delta\eta$, 
\begin{equation}
\delta\phi\lr{x,\eta,t}=\phi_{\ast}\lr{x,\eta,t}-\phi\lr{x,\eta,t}=\frac{\omega}{2}\int_{0}^{x}\delta\eta\,ds-\phi_{y}\delta\eta.\label{dphieta}
\end{equation}
On the other hand, by applying boundary conditions, the variation of the Hamiltonian corresponding to the change $\delta\eta$ can be evaluated to be
\begin{equation}
\delta H=\int_{0}^{L}\left\{\delta\phi\,\eta_{t}+\delta\eta\left[\frac{\abs{\nabla\phi}^2}{2}-\omega\eta \phi_{x}+g\eta+\frac{\omega^2}{2}\eta^2-\alpha\frac{\p}{\p x}\lr{\frac{\eta_{x}}{\sqrt{1+\eta_{x}^2}}}\right]\right\}_{y=\eta}\,dx.\label{dHeta}
\end{equation}
Hence, substituting equations (\ref{dphieta}) and (\ref{eta_x}) into 
(\ref{dHeta}) and integrating by parts, one obtains
\begin{equation}
\frac{\delta H}{\delta\eta}=-\frac{\omega}{2}\lr{\frac{\omega}{2}\eta^2-\chi}-\phi_{y}\eta_{t}+\frac{\abs{\nabla\phi}^2}{2}-\omega\eta\phi_{x}+g\eta+\frac{\omega^2}{2}\eta^2-\alpha\frac{\p}{\p x}\lr{\frac{\eta_{x}}{\sqrt{1+\eta_{x}^2}}}.\label{dHeta2}
\end{equation}
Here, all quantities are evaluated at $y=\eta$.
Finally, at $y=\eta$, we have
\begin{equation}
\zeta_{t}=\vartheta_{t}-\frac{\omega}{2}\int_{0}^{x}\eta_{t}\,ds=
\phi_{t}+\phi_{y}\eta_{t}-\frac{\omega}{2}\lr{\frac{\omega}{2}\eta^2-\chi}
+\frac{\omega}{2}\lr{\frac{\omega}{2}\eta^2\lr{0,t}-\chi\lr{0,t}}.\label{zetat_}
\end{equation}
In the last passage we used equation (\ref{eta_x}). 
Recalling that $\tilde{\chi}\lr{0,t}=\chi\lr{0,t}-\frac{\omega}{2}\eta^2\lr{0,t}=0$ and substituting (\ref{ST}) with $f=0$ in  equation (\ref{dHeta2}), from equation (\ref{zetat_}) we arrive at
\begin{equation}
\zeta_{t}=-\frac{\delta H}{\delta \eta}=-\frac{\abs{\nabla\phi}^2}{2}-g\eta+\alpha\frac{\p}{\p x}\lr{\frac{\eta_{x}}{\sqrt{1+\eta_{x}^2}}}+\phi_{y}\eta_{t}+\omega\eta\phi_{x}-\frac{\omega}{2}\chi-\frac{\omega^2}{4}\eta^2.\label{zetat_2}
\end{equation}
In light of equations (\ref{etat_2}) and (\ref{zetat_2}), we have thus shown that the water wave system (\ref{WWCVB}) can be
cast in canonical Hamiltonian form in terms of the variables $(\zeta,\eta)$.


\subsection{Translational symmetry}
The remaining task is to show that traveling wave solutions of system (\ref{WWCVB}) exhibit translational symmetry. 
A traveling wave solution is defined by the property that it 
can be written as a function of $x'=x-c\,t$, i.e. 
\begin{equation}
\zeta=\zeta\lr{x'},~~~~\eta=\eta\lr{x'}.
\end{equation}
That such solutions, if they exist, are symmetric under translation
can be deduced by the invariance of the action $\mc{S}=\int_{0}^{t}\mc{L}\,dt'$, where $\mc{L}$ is the Lagrangian of the system, under the exchange
$\zeta\lr{x'}\rightarrow \zeta\lr{x'+\xi}$ and $\eta\lr{x'}\rightarrow \eta\lr{x'+\xi}$ for any choice of the displacement $\xi\in\mathbb{R}$.  
Since $\zeta$ acts as a canonical momentum, the Lagrangian $\mc{L}$ is related to the Hamiltonian $H$ according to
\begin{equation} 
\mc{L}=\int_{0}^{L}\zeta\eta_{t}\,dx-H.\label{L1}
\end{equation}
The translational symmetry associated to traveling wave solutions 
implies the existence of a conservation law in accordance with Noether's theorem. 
The conserved quantity is the total momentum in the $x$-direction,
\begin{equation}
M_{x}=\int_{0}^{L}\lr{\int_{-\infty}^{\eta}v_{x}\,dy}\,dx.
\end{equation} 
It can be easily verified that, for traveling wave solutions, $M_{x}$ is constant.

In conclusion, since we have shown that system (\ref{WWCVB}) can be cast in canonical Hamiltonian form, and that traveling wave solutions exhibit translational symmetry, the result proved in sections 3 and 4 applies:
exchange of stability will occur at those wave speeds where energy becomes stationary. Notice that this remains true for traveling waves without reflectional symmetry.

\section{Concluding remarks}

In this paper, the superharmonic stability 
of traveling wave solutions in canonical Hamiltonian systems was examined.
By a linear stability analysis, it was shown that, 
if the traveling wave exhibits translational symmetry,
exchange of stability occurs when energy is stationary with respect to wave speed. 

This result generalizes the analysis of \citet{Saffman85}, who
considered wave profiles with reflectional symmetry. 
Since the present calculations are independent of the specific
physical system under consideration, 
they reflect a general property of Hamiltonian dynamics.
The argument is thus expected to apply to water wave systems 
that possess a proper Hamiltonian structure,
even in the presence of reflectionally asymmetric wave profiles.
In particular, in light of the canonical Hamiltonian formulation
of water waves with constant vorticity due to \citet{Wahlen07}, 
Tanaka's instability (\citet{Tanaka83}) should occur
in water waves with constant vorticity, as discussed in section 5. 
This has been confirmed numerically by \citet{Murashige19}
for wave profiles endowed with reflectional symmetry.

We remark that our result applies to general (noncanonical) Hamiltonian systems \citep{Morrison98} if a suitable change of variables can be found so that the transformed Poisson operator is in `Darboux' form \citep{Littlejohn82}, 
i.e. the system can be described by a set of canonically conjugated variables
plus a given number of constants of motion (Casimir invariants).
This is the case, for example, of the ideal Euler equations, which can be cast in Darboux form in terms of Clebsch parameters \citep{Yoshida17}. 

\section{Acknowledgments}

\noindent The research of N. S. was supported by JSPS KAKENHI Grant No. 18J01729, 
and that of M. Y. by JSPS KAKENHI Grant No. 17H02860.

\end{normalsize}

\end{document}